\documentclass[aps,prl,twocolumn,showpacs,superscriptaddress]{revtex4}
\usepackage{graphicx,amssymb,amsmath,bbm,amsfonts}
\newcommand{\sab}{{\scriptstyle AB}}
\newcommand{\smb}{{\scriptstyle B}}
\newcommand{\smd}{{\scriptstyle D}}
\newcommand{\sma}{{\scriptstyle A}}

\newcommand{\cp}{{\cal P}}
\newcommand{\cc}{{\cal C}}
\newcommand{\ocp}{{\cal \overline P}}
\newcommand{\ovarrho}{{\overline \varrho}}
\newcommand{\oalpha}{{\overline \alpha}}
\newcommand{\iid}{\mathbb{I}}

\newcommand{\ie}{{\it{i.e.}}}
\newcommand{\refeq}[1]{Eq.~(\ref{#1})}
\newcommand{\Tr}{\mathrm{ Tr }}
\newcommand{\ket}[1]{\left| #1 \right\rangle}
\newcommand{\bra}[1]{\left\langle #1 \right|}
\newcommand{\etal}{{\it{et al.}}}
\begin{document}
\title{Non-classicality criteria from phase-space representations 
and information-theoretical constraints are maximally inequivalent}
\author{Alessandro Ferraro}
\affiliation{Department of Physics and Astronomy, University 
College London, Gower Street, London WC1E 6BT, UK}
\author{Matteo G. A. Paris}\email{matteo.paris@fisica.unimi.it}
\affiliation{Dipartimento di Fisica dell'Universit\`a degli Studi di 
Milano, I-20133 Milano, Italy.}
\date{\today}
\begin{abstract}
We consider two celebrated criteria for defining the non-classicality of
bipartite bosonic quantum systems, the first stemming from information
theoretic concepts and the second from physical constraints on the
quantum phase-space. Consequently, two sets of allegedly classical 
states are singled out: i) the set $\cal{C}$
composed of the so called classical-classical (CC) states---separable states
that are locally distinguishable and do not possess quantum discord; ii)
the set $\cal{P}$ of states endowed with a positive P-representation 
(P-classical states)---{mixture of Glauber coherent states 
that, e.g., fail to show
negativity of their Wigner function}. By showing that $\cal{C}$ and
$\cal{P}$ are almost disjoint, we prove that the two defining criteria
are maximally inequivalent.{ Thus, the notions of classicality that they put
forward are radically different.}
In particular, generic CC states show quantumness in their P-representation
and, viceversa, almost all P-classical states have positive quantum
discord, hence are not CC. This inequivalence is
further elucidated considering different applications of P-classical and
CC states. Our results suggest that there are other
quantum correlations in nature than those revealed by entanglement and
quantum discord.  
\end{abstract}
\maketitle
The question of whether a quantum system exhibits a behaviour without
classical analogue has been of interest since the early days of quantum
mechanics. Considering bosonic systems, a major framework for attacking
this question has been established more then half a century ago,
stemming from the notions of quantum phase-space and quasi-probability
distributions \cite{Wig32,Gla63}. There, physical constraints
expressing classical behaviour impose criteria of non-classicality that
have been experimentally tested in a variety of quantum systems
\cite{Smi93,Lei96,Hof09}. On the other hand, in the last two decades
non-classical correlations have been the subject of a renewed interest,
mainly due to the general belief that they are a fundamental resource
for quantum information processing. Within this perspective,
a different approach to non-classicality have emerged, which bases its
ground on the information-theoretic aspects of quantum correlations. In
particular, rigorous criteria to define non-classicality of correlations
have been put forward \cite{Wer89,OZ01,HV01,PHH}, giving rise to well
established concepts like entanglement or quantum discord.  
\par
Here we compare these two approaches, investigating in particular
whether physical constraints emerging from the former can bring new
insight in the assessment of quantum correlations beyond the purely
information-theoretic aspects of the latter. We have found that this is
indeed the case: the notion of non-classical correlations springing from
physical considerations on the quantum phase-space is inequivalent to
that emerging from information-theoretic arguments. In a sense that will
be specified in the following, these two notions of non-classicality are
maximally inequivalent. This, in particular, suggests that there are other
quantum correlations in nature than those revealed by entanglement and
quantum discord.
\par
\textit{Non-classicality in the phase-space}---The uncertainty relations 
make the notion of phase-space in quantum
mechanics problematic. {Following the seminal investigations of Wigner \cite{Wig32}, 
an abundance of quantum mechanical phase-space quasi-distributions were introduced, }
ranging from the Husimi function to the Glauber-Sudarshan P-function \cite{HOSW84}. 
Besides the fundamental aspect, investigations on
quasi-distributions boosted the development of efficient theoretical
tools in various fields of modern physics, e.g. quantum optics and
quantum chemistry \cite{HOSW84,qchem}. These functions cannot, however,
be interpreted as probability distributions over a classical phase-space
because for some quantum states they may be negative or singular.
{Consistently, it is commonly accepted that such features 
underpin a good notion of nonclassicality.}
Supporting this interpretation, fundamental links
between quasi-probability functions and the notions of nonlocality
\cite{BW99} and contextuality \cite{Spe08} have been recognised. 
\par
{In this framework, possibly the most accepted definition of
non-classicality has been introduced by Glauber in terms of the
P-function \cite{Gla63}.} 
For concreteness, let us consider the Hilbert
space ${\cal H}={\cal H}_A \otimes {\cal H}_B$ of a bipartite system
made of two modes $a$ and $b$ of a bosonic field
($[a,a^\dagger]=[b,b^\dagger]=1$). Considering $\alpha, \beta \in
{\mathbb C}$, let us denote with $|\alpha\rangle$ and $|\beta\rangle$
the Glauber coherent states of the systems, that is the eigenstates of
the annihilation operators ($a|\alpha\rangle=\alpha|\alpha\rangle$ and
$b|\beta\rangle=\beta|\beta\rangle$). Any state $\varrho$ of the system
can be expressed in terms of a diagonal mixture of coherent states:
$\varrho = \int\!\!\!\int d^2\alpha\, d^2\beta\:
P(\alpha,\beta)\:|\alpha\rangle\langle\alpha | \otimes |
\beta\rangle\langle \beta|$ where $P(\alpha,\beta)$ is the 
P-function of $\varrho$. When the P-function is a well-behaved
probability density function, then $\varrho$ can be expressed as a
statistical mixture of coherent states \cite{Bon66}. 
Thus, we have
the following classicality criterion: 
\par
\textit{Criterion P (P-classical states).} A state of a bipartite bosonic 
system is \textit{P-classical} if it can be written as
\begin{align}
\varrho_p = \int\!\!\!\int_{\mathbb C} d^2\alpha\, d^2\beta\: P(\alpha,\beta)\:
|\alpha\rangle\langle\alpha | \otimes | \beta\rangle\langle \beta|\,, 
\label{Pc}
\end{align}
where $P(\alpha,\beta)$ is a positive, non-singular, and normalised function.
{This Criterion represents the most conservative notion of
non-classicality in the quasi-probability setting, since when the
P-function is well-behaved so are all other quasi-probabilities.}
The success of using quasi-probabilities to characterise the quantumness 
of a state or its space-time correlations \cite{WM08} is also, loosely speaking, 
related to their ability to capture the difficulty in generating and manipulating quantum
states. In particular in quantum optics, the easiest states to generate
in a lab are coherent and thermal states, characterized by a
well-behaved P-function. On the other hand squeezed, photon-subtracted,
photon-added, and number states, characterized by increasingly
ill-behaved P-functions, happen to be much more difficult to generate.
In this sense, the P-function captures the physical constraints of
producing increasingly-more-quantum states.  Notice, however, that
different coherent states are not orthogonal, hence even when
$P(\alpha,\beta)$ behaves like a true probability density, it does not
describe probabilities of mutually exclusive events. 
\par
\textit{Nonclassicality and information theory}---The first rigorous
attempt to address the classification of quantum correlations from an
information theoretical viewpoint was pioneered by Werner \cite{Wer89}, who put on
firm basis the elusive concept of quantum entanglement
\cite{Hor09,Guh09}. A state of a bipartite system is called entangled if
it cannot be written as follows:  
\begin{align} \label{separable}
\varrho_\sab = \sum p_k \sigma_{\sma k}\otimes\sigma_{\smb k},
\end{align}
where $\sigma_{\sma k}$ and $\sigma_{\smb k}$ are generic density
matrices describing the states of the two subsystems. The definition
above has an immediate operational interpretation: Unentangled
(separable) states can be prepared by local operations and classical
communication between the two parties. One might have thought that such
classical information exchange cannot bring any quantum character to
the correlations in the state. In this sense separability has often been
regarded as a synonymous of classicality in this information theoretical
framework.  
\par
On the other hand, as it has been extensively discussed in the last
decade \cite{OZ01,HV01,Luo08,Ale10,Maz10}, this may not be the case.
An entropic measure of correlations---quantum discord---has been
introduced as the mismatch between the quantum analogues of two classically
equivalent expressions of the mutual information. For pure entangled states, 
quantum discord coincides with
the entropy of entanglement. However, quantum discord can be different
from zero also for (mixed) {separable states}. In other words,
classical communication can give rise to quantum correlations.  This can
be understood by considering that the states $\sigma_{\sma k}$ and
$\sigma_{\smb k}$ in \refeq{separable} may be physically indistinguishable, 
and thus not all the information
about them can be locally retrieved. This phenomenon has no
classical counterpart, thus accounting for the quantumness of the
correlations in separable state with positive discord.
{Few explicit formulas have been derived for the quantum discord of some
states \cite{Luo08,Gio10,Lu11}, 
and more general entropic measures of nonclassical  correlations have been also discussed \cite{Ani11}.
Discord finds an operational meaning in terms of quantum state merging 
\cite{Cav11}, and its role has been studied in quantum information processing
with mixed states, where there are computational and communication
tasks which are seemingly impossible to achieve classically, and yet can
be attained using little or no entanglement \cite{Kni98,Dat08,Div04}. } 
More recently, monogamy properties of discord have been investigated \cite{glg11}, and
it has been shown that quantum correlations in separable states may be
activated into distillable entanglement \cite{Pia11}.  Discord is also
related to the minimum entanglement generated between system and
apparatus in a partial measurement process \cite{Str11}.
\par
Remarkably, even states with zero discord can show non-classical
correlations. In order to see this effect in details let us recall that
discord is asymmetric in the two modes and that a bipartite state with
zero A-discord can be written in the form $\varrho_{\sab}=\sum_k p_k
|\theta_k \rangle\langle \theta_k | \otimes \sigma_{\smb k}\,,$ where
the $|\theta_k\rangle$'s form an orthonormal basis and the $\sigma_{\smb
k}$'s are a set of generic non-orthogonal states.  These states---dubbed
quantum-classical states---cannot be cloned locally (locally
broadcasted), despite having zero discord \cite{PHH}. This security
against local broadcasting is not featured by any correlated state of a
classical system, thus revealing the quantumness of this type of zero
discord states. 
The set of states that can be locally broadcasted has been shown to be
equivalent to a set of states called classical-classical (CC) \cite{PHH}. Any member of
such set can be written as
\begin{align}\label{cc}
\varrho_{AB}=\sum_{ks} p_{ks} |\theta_k
\rangle\langle \theta_k | \otimes 
|\eta_s
\rangle\langle \eta_s |,
\end{align}
where $|\theta_k\rangle$ and $|\eta_s\rangle$ are basis for the Hilbert
spaces of the two subsystems. These states are now commonly regarded as purely
classical correlated states \cite{Pia11}. The reason for this is based on
information theoretic arguments. All the information encoded in a
CC state can be locally retrieved and stored in a classical register. 
Indeed, states  appearing in (\ref{cc}) are perfectly 
distinguishable by local quantum measurements. 
In this sense, CC states simply accommodate the joint probability
$p_{ks}$ in a quantum formalism{, thus putting forward the most
conservative notion of non-classicality in an information-theoretical
setting}.  However, we will show in the following that also this class
of states can exhibit quantum correlations that cannot be featured by
systems that admit a classical description in the quantum phase-space.  
\par
Definition (\ref{cc}) was introduced in the context of
finite-dimensional systems and it needs to be slightly generalized in
order to fully take into account some subtleties of bosonic systems for
which there exists basis that are unitarily inequivalent
\cite{footnote1}. Considering $x,y \in {\mathbb R}$, let us denote with
$|x\rangle$ and $|y\rangle$ two generic basis of $A$ and $B$
respectively. We introduce the following classical criterion: \par
\textit{Criterion C (classical-classical states).} A state of a
bipartite bosonic system is CC if it can be
written as 
\begin{align}
\varrho_c = \int\!\!\!\int_{\mathbb R} dx dy \: F(x,y)\:
|x\rangle\langle x | \otimes | y\rangle\langle y|
\label{ccc}
\end{align}
and $F(x,y)$ is a positive, non-singular, and normalised function.
Notice that, in general, the joint probability distribution $F(x,y)$
spans over a continuous set. Clearly, one recovers Eq. (\ref{cc}) if
$F(x,y)$ is non-zero only over a discrete set. 
\par 
\textit{Number correlated states}---In the following we show that the
foregoing criteria of non-classicality are maximally inequivalent.
However, before proceeding with a formal proof, let us discuss a
specific example. Consider the two-mode P-classical states introduced in
\refeq{Pc}, and define the observable $O_\smd=a^\dag a - b^\dag b$,
which detects the difference between the number of quanta of the two
modes.
Since for coherent states $\langle z| a^\dag a | z\rangle = |z|^2$ and 
$\langle z| (a^\dag a)^2 | z\rangle = |z|^4 + |z|^2${, for any P-classical
state (different from the vacuum) we have} 
\begin{align}\label{diff}
\Delta O^2_\smd = |\alpha_0|^2 + |\beta_0|^2 + \mathrm{Tr}\, C \geq 
|\alpha_0|^2 + |\beta_0|^2 >0
\end{align}
being $\alpha_0$, $\beta_0$ and $C$ the mean values and the 
covariance matrix of $P(\alpha,\beta)$ respectively. 
The observable $O_\smd$ detects correlations between the number of quanta
in the two modes. The above inequality captures the intuition behind the 
idea that the behaviour of a classical state should be that of a mixture 
of coherent states: each mode has a fluctuating number
of quanta and the difference should fluctuate accordingly. In other
words, for a classical two-mode system the amount of intensity 
correlations between two modes is bounded.
\par
Let us now consider the two modes prepared in the state $\varrho_{nc} =
\sum_n p_n |n\rangle\langle n| \otimes |n\rangle\langle n|$, where 
$a^\dagger a \ket{n}=n\ket{n}$. This is the
state generated by, say, a pair of machine guns, each producing a random
but equal number of bullets $n$ according to the distribution $p_n$. The
state $\varrho_{nc}$ is separable and, according to the terminology
introduced above, CC. Yet it shows perfect
correlations in the number of quanta.  Actually, the product states
$|n\rangle\langle n| \otimes |n\rangle\langle n|$ are the projectors
over the degenerate eigenspace of $O_\smd$ with eigenvalue zero. In
other words, for any choice of the distribution $\{p_n\}$ we have
$\Delta O^2_\smd=0$ for $\varrho_{nc}$, which in turn violates the
inequality (\ref{diff}). Thus the family of number correlated states
$\varrho_{nc}$ gives an example of states that obey \textit{Criterion C}
while violating \textit{Criterion P}.  We will now proceed to
prove that the two criterion are not only inequivalent, but that their
inequivalence is maximal. Specifically we will show that {generic}
states obeying \textit{Criterion P} violates \textit{Criterion C} and
vice-versa.  
\par
\textit{{Generic} P-classical states are not CC}---Consider the 
following property of any CCstate
(necessary condition for CC states): any two states of
system $A$ conditioned to a measurement on $B$ commute. This can be seen
by considering the definition in \refeq{ccc} and applying any POVM on
$B$. It immediately follows that any state of $A$ conditioned on any outcome at $B$
will remain diagonal in the original basis. Thus, all possible
conditioned states of $A$ will mutually commute. 
\par 
Consider now a generic
P-classical state and the following two convenient conditioned states of $A$: 
$\varrho_{A}=\Tr_B{[\varrho_p]}=\int\!\! d^2\alpha\: P(\alpha)\:
|\alpha\rangle\langle\alpha |$, and 
$\varrho_{0}=\Tr_B{[\varrho_p \ket{0}\bra{0}]}=\int\!\! d^2\alpha\: P_0(\alpha)\:
|\alpha\rangle\langle\alpha |$,
where $P(\alpha)=\int d^2\beta\: P(\alpha,\beta)$, $P_0(\alpha)=\int
d^2\beta\: P(\alpha,\beta) e^{-|\beta|^2}$, and $\ket{0}\bra{0}$ is the
vacuum. Calculating the commutator between the above states and
evaluating it on the vacuum, one has 
\begin{align}\label{comm_0}
\bra{0}[\varrho_{A},\varrho_{0}]\ket{0}=\int &d^2\alpha\, d^2\alpha'\: 
P(\alpha)P_0(\alpha') \nonumber \\ 
&e^{-|\alpha|^2}e^{-|\alpha'|^2}(e^{\alpha\overline{\alpha'}}-c.c.).
\end{align}
Imposing that the commutator above is identical to zero yields the following nontrivial constraint on the P-function
$P(\alpha,\beta)$: 
$
\int d^2\alpha\, d^2\alpha' d^2\beta\, d^2\beta'\:
P(\alpha,\beta)P(\alpha',\beta')  \times  \\ 
e^{-|\alpha|^2}e^{-|\alpha'|^2}e^{-|\beta'|^2}
(e^{\alpha\overline{\alpha'}}-c.c.)=0 \notag\,.
$
A generic (well-behavied) P-function does not satisfy the above
constraint. This, in turn, implies that \textit{almost all P-classical states
are not CC}. Equivalently, generic P-classical states violate \textit{Criterion C}. 
Notice that the proof works as well for
$A$-discord states, thus showing that almost all P-classical states have
positive discord.
\par
\textit{{Generic} CC states are not P-classical}---We
first need to show that the set $\cp$ of single mode P-classical states
is nowhere dense in the bosonic space. By definition, $\cp$ is nowhere
dense if its closure ${\cal \overline P}$ has no interior points.
Denoting by $\partial \cp$ the frontier of ${\cal P}$ (namely, the set
of its accumulation points), one has that  ${\cal \overline P}=\cp \cup
\partial \cp$. The P-function of any operator $\delta\in \partial \cp$
must be positive since it is the limit of positive functions. In
addition, it cannot be singular everywhere in the phase space, given
that it is the limit of normalizable functions. As a consequence any
operator $\ovarrho\in\ocp$ is such that its P-function is positive and
not everywhere singular. Let us now show that no $\ovarrho$
can be an interior point of $\ocp$. First, given any $\ovarrho$ denote
by $\oalpha$ a point in the phase space where the P-function of
$\ovarrho$ is non-singular (\ie, $P_{\ovarrho}(\oalpha)<\infty$). Then
define a convenient perturbation of $\ovarrho$: $\varrho=(1-\epsilon)
\ovarrho+\epsilon D(\oalpha) \varrho_1 D^\dagger(\oalpha)$,  
where $0<\epsilon<1$, $D(\oalpha)=\exp[\oalpha a^\dagger - 
\oalpha^* a]$ is the displacement operator, and $\varrho_1=
\ket{1}\bra{1}$ is a single excitation state. One has that the P-function 
of $\varrho$ is given by: $P_\varrho(\alpha)=(1-\epsilon)P_\ovarrho(\alpha)
+\epsilon P_{\varrho_1} (\alpha-\oalpha)$. 
Since the P-function of the single excitation state is negative and
singular at the origin, one has that $P_\varrho(\alpha)$ is non-positive
(and singular in $\oalpha$). For what shown above, this means that (for
any $\epsilon$) $\varrho\notin\ocp$, hence $\ovarrho$ is not an interior
point of $\ocp$. Since this holds true for any $\ovarrho$, one has that
$\ocp$ has no interior points. As a consequence $P$ is nowhere dense in
the space of single mode bosonic systems.   
\par 
{Consider now the set $\cp_2$ of two-mode P-classical states. Based on
the above considerations one can show that P-classical states $\varrho_p
\in  \cp_2$ are nowhere dense in the set $\cc$ of CC states. First,
recall that the partial trace of any P-classical state is a P-classical
state (necessary condition for P-classical states). This implies that
the partial trace of any $\ovarrho_p \in  \ocp_2$ must have a
non-negative P-function. Then, using the same arguments as above
(technical details are omitted), one can build a CC state $\varrho'$
that, despite being an infinitesimal perturbation of $\ovarrho_p$, does
not belong to $\ocp_2$. This implies that $\ocp_2$ has no interior point
in $\cc$, hence \textit{P-classical states are nowhere dense in the set
of CC states}. Equivalently, generic CC states violate \textit{Criterion
P}.} \par
\textit{Discussion}---{The foregoing arguments show that the set of
states simultaneously obeying \textit{Criteria P} and \textit{C} is
negligible, both in a metrical and topological sense \cite{note}. In other words,
the two criteria considered here put forward two radically different
notions of classicality of correlations. \textit{Criterion C} looks at
the correlations between the information of A and B, as encoded in their
states and
regardless the quantumness of the states themselves,}
whereas \textit{Criterion P} takes into account physical constraints on those
as well. Referring to the example of number correlated states $\varrho_{nc}$: 
creating Fock states with the same
number of quanta does correspond to establishing quantum correlations
between the modes, irrespectively from the fact that the information
needed to perform this action may be of purely classical (local) origin.
It has been often argued that a suitable quantity to reveal quantum
correlations in bipartite systems, {beyond the presence of
entanglement, should be related to the joint information carried by the
state. For example, quantum discord focus on this and can be used to assess states
for application in quantum communication. On the other hand, from a
fundamental physical point of view, discord (and information-theoretical 
quantities more in general) appears unable to account
for the very physical constraints involved in the establishment of correlations. 
Ultimately, this means that allegedly} classical
correlations established between systems prepared in states with no
classical analogue are quantum in nature. 
\par
Operationally, the fact that P-classical states violate
\textit{Criterion C} allows to use them {as an experimentally cheap resource }
in communication protocols that
require security against local broadcasting. On the other hand, the
nonclassicality of CC states like $\varrho_{nc}$ may find an operational
characterization in terms of conditional measurements. Consider a
generic bipartite state and perform a measurement described by the POVM
$\{\Pi_x\}$ on one mode, say mode $1$. If the state is P-classical then
the P-function of the conditional state $\varrho_{px} = \hbox{Tr}_1
[\varrho_p\, \Pi_x \otimes \iid]/p_x$ may be written as
$$
P(\beta)= \frac1{p_x} \int\! d^2\alpha\,
P(\alpha,\beta)\,\langle\alpha|\Pi_x|\alpha\rangle\,.
$$
This is a well behaved probability density function, and thus the
state $\varrho_{px}$ is classical. In other words, only states 
violating \textit{Criterion P} may lead to the conditional generation 
of genuine quantum states with no classical analogue \cite{fer04,bon07}. 
\par
\textit{Conclusions}---
{In the last two decades the fruitful exchange of notions between
information science and quantum physics led to the emergence of
radically new concepts and applications. The slogan {\em information is
physical} \cite{lan93} has become increasingly popular, emphasizing the
role of physical constraints in quantum information processing
\cite{dvl98}. Our results reinforce this position, however also present
an unusual case in which the information-theoretical and physical
perspectives appear fundamentally conflicting. Specifically, by
addressing the notion of non-classicality as it emerges from physical
considerations, we have shown that there exist other genuinely quantum
correlations than those revealed by information-theoretic arguments.
This indicates that the slogan should be complemented by a second part
illustrating that information-theoretic considerations cannot substitute
physical constraints, thus suggesting that {\em information is physical,
and physics is not merely information}.}
\par
MGAP thanks Paolo Giorda, Sabrina Maniscalco, Kavan Modi and Jyrki Piilo 
for interesting discussions. This work has been supported by MIUR (FIRB
LiCHIS-RBFR10YQ3H) and by
the Finnish Cultural Foundation.


\begin{thebibliography}{9}
\bibitem{footnote1} Namely, basis that cannot be inter-converted via a unitary transformation, such as the position $|q\rangle$ [with $\hat{q}|q\rangle=q|q\rangle$ and $\hat{q}=(a+a^\dagger)/\sqrt{2}$] and number $|n\rangle$ [with $a^\dagger a|n\rangle=n|n\rangle$] basis.
\end{thebibliography}
\end{document}